\documentclass[aps,prd,preprint,superscriptaddress,amsmath,amssymb,showpacs]{revtex4-1}
\usepackage{dcolumn}
\usepackage{graphicx}
\usepackage{float}
\usepackage{physics}
\usepackage[colorlinks=true,allcolors=blue]{hyperref}

\begin{document}

\title{Gravitational waves from holographic first-order QCD phase transition with magnetic field}

\author{Man-Man Sun}
\email{sunmm@zknu.edu.cn}
\affiliation{School of Physics and Telecommunications Engineering, Zhoukou Normal University, Zhoukou 466001, China}

\author{Man-Li Tian }
\email{20252018@zknu.edu.cn}
\affiliation{University Clinic, Zhoukou Normal University, Zhoukou 466001, China}

\author{Zhou-Run Zhu }
\email{zhuzhourun@zknu.edu.cn}
\affiliation{School of Physics and Telecommunications Engineering, Zhoukou Normal University, Zhoukou 466001, China}

\date{\today}

\begin{abstract}
In this paper, we investigate the generation of gravitational waves (GWs) from a first-order QCD confinement-deconfinement phase transition under external magnetic field from holography. We analyze the GWs spectra across both hard wall and soft wall models for Jouguet detonations and non-runaway scenarios. Our results indicate that increasing the magnetic field shifts the spectral peak to lower frequencies. The predicted GWs signals are potentially detectable by observatories such as IPTA, SKA, BBO and NANOGrav. Decomposing the spectra reveals that sound waves typically dominate the signal around the peak frequency, bubble collisions prevail at spectral extremities, and the contribution from MHD turbulence is significant only for non-runaway bubble scenarios at high frequencies. This work suggests that magnetized QCD phase transitions are viable cosmological sources for observable GW backgrounds, offering a potential pathway to constrain primordial magnetic fields through future PTA observations.
\end{abstract}

\maketitle

\section{Introduction}\label{sec:01_intro}

The detection of gravitational waves (GWs) signal at the LIGO \cite{LIGOScientific:2016aoc} has opened an unprecedented probe into strong-field gravity and cosmic evolution. Concurrently, the pursuit of a stochastic gravitational wave background (SGWB) in the nanohertz regime has progressed significantly, driven by pulsar timing array (PTA) collaborations \cite{NANOGrav:2020bcs,NANOGrav:2023gor,NANOGrav:2023hde,Moore:2014lga,Ruan:2018tsw}. The sources of gravitational waves can originate from astrophysical or cosmological phenomena \cite{Cai:2017cbj}. Among the plethora of potential cosmological sources for this signal, first-order phase transitions in the early universe are particularly compelling candidates \cite{Caprini:2015zlo,Hindmarsh:2013xza,Ellis:2018mja,Schwaller:2015tja}.

A prime candidate for such a cosmological first-order phase transition is the QCD confinement-deconfinement transition \cite{Kajantie:1996mn,Stephanov:1996ki,Fodor:2004nz}. As the Universe cools through this scale, a violent first-order transition involving bubble nucleation, expansion, and collision can generate a significant SGWB via three primary mechanisms: bubble wall collisions, sound waves in the relativistic plasma, and magnetohydrodynamic (MHD) turbulence \cite{Kosowsky:1991ua,Kosowsky:1992vn,Kamionkowski:1993fg,Caprini:2007xq,Hindmarsh:2015qta,Giblin:2014qia,Caprini:2009yp,Kahniashvili:2008pf,Kisslinger:2015hua,Bigazzi:2021ucw,Yamada:2025hfs}. A first-order phase transition is characterized by the coexistence of two local minima of the free energy across a finite temperature interval. Within this regime, the scalar field may transition into the new phase via quantum tunneling or thermal activation, initiating the nucleation of bubbles in the symmetry-broken phase. These bubbles subsequently expand, merge, and collide in a highly dynamic and inhomogeneous manner. Such a violent spacetime disturbance serves as a potent source of gravitational wave radiation. The characteristic frequency of this signal naturally aligns with the PTA observational window \cite{Caprini:2010xv,Anand:2017kar,Chen:2019xse}.

One critical factor is the potential influence of magnetic field, which produced in the heavy ion collisions and the early Universe \cite{Grasso:2000wj,Subramanian:2015lua,Skokov:2009qp}. The impact of strong magnetic fields on the QCD phase structure is profound and non-trivial. Lattice QCD simulations reveal the phenomenon of inverse magnetic catalysis (IMC) \cite{Bali:2011qj,Bali:2012zg,Ilgenfritz:2013ara}. This stands in contrast to magnetic catalysis (MC) predicted by many effective models and early holographic approaches \cite{Gusynin:1994re,Miransky:2002rp,Gatto:2010pt,Preis:2010cq}. Understanding this interplay is essential not only for mapping the QCD phase diagram under extreme conditions \cite{DElia:2011koc,Fukushima:2012xw,Farias:2014eca,Mamo:2015dea,Dudal:2015wfn} but also for accurately predicting the resulting gravitational waves signature. The magnetic field modifies the thermodynamics, transition dynamics, and plasma properties, thereby imprinting distinct features on the SGWB spectrum \cite{Mizher:2010zb,Han:2016uhh}.

Recent years, the investigations of gravitational waves by using the AdS/CFT correspondence \cite{Witten:1998qj,Gubser:1998bc,Maldacena:1997re} have attracted much attention. The authors of Ref. \cite{Ahmadvand:2017xrw,Li:2021qer,Chen:2023bms} have studied the GWs within the AdS/QCD. Further analyses have discussed the effect of finite chemical potential on the GWs under varied bubble wall velocities \cite{Ahmadvand:2017tue}. The influence of coupling corrections on the GWs spectrum has been systematically investigated in \cite{Rezapour:2020mvi,Rezapour:2022iqq}. The GWs production associated with the QCD phase transition also have been studied in the holographic QCD models \cite{Chen:2017cyc,Ares:2020lbt,Cai:2022omk,He:2022amv,He:2023ado,Chen:2022cgj}. Additionally, the impact of gluon condensate and hyperscaling violation on the GWs have been studied in \cite{Zhu:2021vkj} and \cite{Zhu:2024zhc} respectively.

In this work, we investigate the stochastic gravitational waves generated by a first-order QCD confinement-deconfinement phase transition in the presence of a strong external magnetic field from holography. We employ the bottom-up holographic framework, building upon the magnetized black hole solutions \cite{DHoker:2009mmn,DHoker:2009ixq} and incorporating a phenomenological dilaton to model confinement \cite{Mamo:2015dea,Dudal:2015wfn}. Within this setup, we explore the effect of magnetic field on the GWs spectrum. Our analysis aims to determine if magnetized QCD phase transitions remain viable cosmological sources for the observed PTA signal and to establish the potential of future PTA observations to constrain the strength and effects of primordial magnetic fields.

The paper is organized as follows. In Sec.~\ref{sec:02}, we discuss the holographic QCD phase transition with magnetic field. In Sec.~\ref{sec:03}, we explore the effect of magnetic field on the gravitational waves spectrum. The conclusion and discussion are given in Sec.~\ref{sec:04}.

\section{Holographic QCD phase transition with magnetic field}\label{sec:02}

In this section, we review the research of Ref. \cite{Dudal:2015wfn,Mamo:2015dea}, which study the confinement-deconfinement phase transition in the presence of external magnetic field from hard wall and soft wall models. The Einstein-Maxwell action is given by \cite{DHoker:2009mmn,DHoker:2009ixq}:
\begin{equation}
S=S_{\text{bulk}}+S_{\text{bndy}},
\label{action}
\end{equation}
where $S_{\text{bulk}}$ is:
\begin{equation}
S_{\text{bulk}}=\frac{1}{16\pi G_5}\int d^5x\sqrt{-g}\left(R-F^{MN}F_{MN}+\frac{12}{L^2}\right),
\label{sbulk}
\end{equation}
with $\sqrt{-g}=\sqrt{-\det g_{\mu\nu}}$. $F_{MN}$ and $R$ denote the electromagnetic field strength and the Ricci scalar respectively. The negative cosmological constant is given by $\Lambda=-\frac{12}{L^2}$, where $L$ represents the AdS radius. In our calculations, we set $L=1$. The 5D Newton constant $G_5 = \dfrac{45\pi L^3}{16(N_c^2-1)}$ is obtained by requiring that the high-temperature limit of the free energy reproduces the Stefan-Boltzmann result, ensuring correct counting of degrees of freedom in the deconfined phase \cite{Dudal:2015wfn}. $N_c =3$ is the number of colors.

$S_{\text{bndy}}$ is the boundary action, which incorporates both the Gibbons-Hawking boundary term and holographic counterterms introduced to cancel UV divergences. The form of $S_{\text{bndy}}$ is \cite{Dudal:2015wfn}
\begin{equation}
S_{\text{bndy}}=\frac{1}{8\pi G_5}\int d^4x\sqrt{-\gamma}\left(K-\frac{3}{L}-\frac{L}{2}F^{\mu\nu}F_{\mu\nu}\left(\ln \frac{r}{L}\right)\right)\Bigg|_{r_\lambda},
\label{sbndy}
\end{equation}
where $r$ is the radial holographic coordinate. $r_\lambda$ is used to regulate the divergence at $r=0$. Meanwhile, $\gamma$ represents the determinant of the induced metric $\gamma_{\mu\nu}$. $K$ denotes the trace of the extrinsic curvature.

The confinement-deconfinement phase transition is dual to Hawking-Page transition between magnetized thermal AdS (confined) and black hole (deconfined) phases. The confinement-deconfinement phase transition occurs when the free energy density difference $\Delta F$ reaches zero. The free energy $F$ can be calculated from the on-shell actions as $S=\beta F$. In the hard wall model, the bulk action is  \cite{Dudal:2015wfn}
\begin{equation}
\label{euclideanbulkaction}
S_{\text{bulk}}=\frac{V_3}{8\pi G_5}\int_0^\beta dt_E\int_{r_\lambda}^{r'}dr\sqrt{g}\left(\frac{4}{L^2}+\frac{2}{3}B^2g^{xx}g^{yy}\right),
\end{equation}
where $V_3$ denotes the volume. It should be mentioned that $r'=R_H$ in black hole case, while $r'=r_0$ in thermal AdS case. The hard-wall IR cutoff $r_0 = 3.096\ \mathrm{GeV}^{-1}$ is determined by matching the mass of the lowest $\rho$ meson \cite{Herzog:2006ra}.

The boundary action is \cite{Dudal:2015wfn}
\begin{equation}
S_{\text{bndy}}=-\frac{V_3}{8\pi G_5}\int_0^\beta dt_E\sqrt{-\gamma}\left(K-\frac{3}{L}-LB^2g^{xx}g^{yy}\left(\ln \frac{r}{L}\right)\right)\Bigg|_{r_\lambda}.
\label{euclideanboundaryaction}
\end{equation}

The action of the black hole (deconfined phase) for the hard wall model is \cite{Dudal:2015wfn}
\begin{eqnarray}
&S_{\text{bh}}&=S_{\text{bulk}}^{\text{bh}}+S_{\text{bndy}}^{\text{bh}}\\
&=&\frac{V_3L^3}{8\pi G_5}\beta\left[-\frac{1}{R_H^4}+\frac{1}{2r_h^4}+\frac{B^2}{3L^2}+\frac{2B^2}{3L^2}\ln\left(\frac{R_H}{r_\lambda}\right) -\frac{B^2}{3L^2}\ln\left(\frac{r_\lambda}{\ell_d}\right) +\frac{B^2}{L^2}\ln\left(\frac{r_\lambda}{L}\right)\right]. \nonumber
\end{eqnarray}

The action of the thermal magnetized AdS geometry (confined phase) for the hard wall model is \cite{Dudal:2015wfn}
\begin{eqnarray}
&S_{\text{th}}&=S_{\text{bulk}}^{\text{th}}+S_{\text{bndy}}^{\text{th}}\nonumber\\
&=&\frac{V_3L^3}{8\pi G_5}\beta\left[-\frac{1}{r_0^4}+\frac{B^2}{3L^2}+\frac{2B^2}{3L^2}\ln\left(\frac{r_0}{r_\lambda}\right)-\frac{B^2}{3L^2}\ln\left(\frac{r_\lambda}{\ell_c}\right)+\frac{B^2}{L^2}\ln\left(\frac{r_\lambda}{L}\right)\right].
\end{eqnarray}

The phase transition takes place when the state of minimum free energy shifts between the two phases. Consequently, we determine the transition temperature by $\Delta S=0$ and the expression of $\Delta S$ is \cite{Dudal:2015wfn}
\begin{eqnarray}\label{88}
\Delta S&=&S_{\text{bh}}-S_{\text{th}}\nonumber\\
&=&\frac{V_3L^3}{8\pi G_5}\beta\left[\frac{-1}{2R_H^4}+\frac{1}{r_0^4}+\frac{B^2}{3L^2}\ln\left(\frac{R_H^3}{\ell_c r_0^2}\right)\right].
\end{eqnarray}

Since the horizon $R_H$ is determined by the temperature $T$ and magnetic field $B$ via the Hawking temperature formula, this equation establishes the functional dependence of the phase transition temperature on the magnetic field.

As noted in \cite{Herzog:2006ra}, studying the Hawking-Page transition within the soft-wall model requires the assumption that the dilaton field $\phi = cr^2$ does not strongly backreact on the metric. The influence of dilaton field on the gravitational dynamics is taken to be negligible. The parameter $c = 0.151\ \mathrm{GeV}^{2}$ is fixed by the linear Regge trajectory of $\rho$ meson radial excitations, directly linking the model to the QCD spectrum \cite{Herzog:2006ra}.

The action for the black hole (deconfined phase) in the soft wall model is \cite{Dudal:2015wfn}
\begin{eqnarray}\label{89}
\begin{split}
S_{\text{bh}}=&\frac{V_3L^3}{8\pi G_5}\beta\left[e^{-cR_H^2}\left(\frac{-1}{R_H^4}+\frac{c}{R_H^2}\right)+\left(\frac{B^2}{3L^2}+c^2\right)\text{Ei}(-cR_H^2)-e^{-cr_\lambda^2}\left(\frac{-1}{r_\lambda^4}+\frac{c}{r_\lambda^2}\right)\right.\\
&-\left.\left(\frac{B^2}{3L^2}+c^2\right)\text{Ei}(-cr_\lambda^2)-\frac{1}{r_\lambda^4}+\frac{1}{2r_h^4}+\frac{B^2}{3L^2}-\frac{1}{3}\frac{B^2}{L^2}\ln \left(\frac{r_\lambda}{\ell_d}\right)+\frac{B^2}{L^2}\ln \left(\frac{r_\lambda}{L}\right)\right].
 \end{split}
\end{eqnarray}

In the soft-wall model, the action of thermal magnetized AdS (confined phase) is \cite{Dudal:2015wfn}
\begin{align}
&S_{\text{th}}=\frac{V_3L^3}{8\pi G_5}\beta\left[-e^{-cr_\lambda^2}\left(\frac{-1}{r_\lambda^4}+\frac{c}{r_\lambda^2}\right)-\left(\frac{B^2}{3L^2}+c^2\right)\text{Ei}(-cr_\lambda^2)-\frac{1}{r_\lambda^4}+\frac{B^2}{3L^2}-\frac{1}{3}\frac{B^2}{L^2}\ln\left(\frac{r_\lambda}{\ell_c}\right) \right.\nonumber\\&\left.+ \frac{B^2}{L^2}\ln\left(\frac{r_\lambda}{L}\right)\right].
\end{align}

Then one can get the action difference of soft wall case \cite{Dudal:2015wfn}
\begin{equation}
\label{a1}
\begin{split}
\Delta S&=\frac{V_3L^3}{8\pi G_5}\beta\left[e^{-cR_H^2}\left(\frac{-1}{R_H^4}+\frac{c}{R_H^2}\right)+\left(\frac{B^2}{3L^2}+c^2\right)\text{Ei}(-cR_H^2)+\frac{1}{2R_H^4}+\frac{1}{3}\frac{B^2}{L^2}\ln\left(\frac{R_H}{\ell_c}\right) \right],
\end{split}
\end{equation}
where $\ell_c = 1.03\ \mathrm{GeV}^{-1}$ appearing in the confined (thermal AdS) phase is constrained in Ref.~\cite{Dudal:2015wfn} by requiring both the absence of naked curvature singularities and consistency with the lattice QCD behavior of the chiral condensate under a magnetic field at zero temperature.

The Hawking temperature is given by \cite{Dudal:2015wfn}
\begin{equation}
\label{hawktem}
T=\frac{1}{4\pi}\bigg|\frac{4}{R_H}-\frac{2}{3}\frac{B^2R_H^3}{L^2}\bigg|.
\end{equation}

\section{Gravitational waves from holographic QCD phase transition with magnetic field}\label{sec:03}

In this section, we explore the GWs from holographic QCD phase transition with magnetic field. A first-order phase transition is characterized by bubble nucleation within a supercooled plasma. The expansion of these bubble walls is powered by the latent heat released from the free energy difference between the symmetric and broken phases. Gravitational waves are produced primarily from the collision of bubbles ($h^{2}\Omega_{env}$) \cite{Huber:2008hg,Jinno:2016vai}, the subsequent sound waves ($h^{2}\Omega_{sw}$) \cite{Hindmarsh:2015qta,Hindmarsh:2017gnf}, and the magnetohydrodynamic (MHD) turbulence generated in the fluid ($h^{2}\Omega_{turb}$) \cite{Caprini:2009yp,Binetruy:2012ze}. Thus, the
total GWs frequency spectrum ($h^{2}\Omega(f)$) is
\begin{equation}\label{a}
h^{2}\Omega(f)=h^{2}\Omega_{env}(f)+h^{2}\Omega_{sw}(f)+h^{2}\Omega_{turb}(f)~,
\end{equation}
where
\begin{equation}\label{b}
\begin{split}
 &h^{2}\Omega_{env}(f)=3.6\times10^{-5}\left(\frac{0.11v_{b}^{3}}{0.42+v_{b}^{2}}\right)\left(\frac{H_{\ast}}{\tau}\right)^{2}\left(\frac{\kappa_1 \alpha}{1+\alpha}\right)^{2}\left(\frac{10}{g_{\ast}}\right)^{\frac{1}{3}}S_{env}(f),\\
&h^{2}\Omega_{sw}(f)=5.7\times10^{-6}\left(\frac{H_{\ast}}{\tau}\right)\left(\frac{\kappa_{2}\alpha}{1+\alpha}\right)^{2}\left(\frac{10}{g_{\ast}}\right)^{\frac{1}{3}}v_{b}S_{sw}(f),\\
&h^{2}\Omega_{turb}(f)=7.2\times10^{-4}\left(\frac{H_{\ast}}{\tau}\right)\left(\frac{\kappa_{3}\alpha}{1+\alpha}\right)^{\frac{3}{2}}\left(\frac{10}{g_{\ast}}\right)^{\frac{1}{3}}v_{b}S_{tu}(f).
\end{split}
\end{equation}

The spectral shapes of gravitational waves are
\begin{equation}\label{c}
\begin{split}
&S_{env}(f)=\frac{3.8\left(\frac{f}{f_{env}}\right)^{2.8}}{1+2.8\left(\frac{f}{f_{env}}\right)^{3.8}},
\\
&S_{sw}(f)=\left(\frac{f}{f_{sw}}\right)^{3}\left(\frac{7}{4+3\left(\frac{f}{f_{sw}}\right)^{2}}\right)^{\frac{7}{2}},
\\
&S_{turb}(f)=\frac{\left(\frac{f}{f_{turb}}\right)^{3}}{\left(1+\frac{f}{f_{turb}}\right)^{\frac{11}{3}}\left(1+\frac{8\pi f}{h_{\ast}}\right)},
\end{split}
\end{equation}
with
\begin{equation}\label{d}
h_{\ast}=11.2\times10^{-9}[\text{Hz}]\left(\frac{T_{\ast}}{100\text{MeV}}\right)\left(\frac{g_{\ast}}{10}\right)^{\frac{1}{6}}.
\end{equation}

The peak frequency of each gravitational waves spectrum is
\begin{equation}\label{e}
\begin{split}
&f_{env}=11.2\times10^{-9}[\text{Hz}]\left(\frac{0.62}{1.8-0.1v_{b}+v_{b}^{2}}\right)\left(\frac{\tau}{H_{\ast}}\right)\left(\frac{T_{\ast}}{100\text{MeV}}\right)\left(\frac{g_{\ast}}{10}\right)^{\frac{1}{6}},
\\
&f_{sw}=12.9\times10^{-9}[\text{Hz}]\left(\frac{1}{v_{b}}\right)\left(\frac{\tau}{H_{\ast}}\right)\left(\frac{T_{\ast}}{100\text{MeV}}\right)\left(\frac{g_{\ast}}{10}\right)^{\frac{1}{6}},
\\
&f_{turb}=18.4\times10^{-9}[\text{Hz}]\left(\frac{1}{v_{b}}\right)\left(\frac{\tau}{H_{\ast}}\right)\left(\frac{T_{\ast}}{100\text{MeV}}\right)\left(\frac{g_{\ast}}{10}\right)^{\frac{1}{6}}.
\end{split}
\end{equation}

In this work, we explore two bubble scenarios: the Jouguet detonations and the non-runaway bubbles \cite{Caprini:2015zlo,Hindmarsh:2015qta,Kamionkowski:1993fg}. For Jouguet detonations, $\kappa_1 = (0.715 \alpha +0.181 \sqrt{\alpha})/(1+ 0.715 \alpha)$, $ \kappa_2 = \sqrt{\alpha}/(0.135 + \sqrt{\alpha +0.98})$, $\kappa_3 = 0.05 \kappa_2$, and $v_b = (\sqrt{1/3} +\sqrt{\alpha^2 +2 \alpha/3})/( 1+ \alpha)$. In non-runaway bubbles case, $\kappa_1 = 0$, $\kappa_2 = \alpha/(0.73 +0.083 \sqrt{\alpha} +\alpha)$, $\kappa_3 = 0.05 \kappa_2$, and $v_b = 0.95$. Here, $\alpha=\frac{\epsilon_{\ast}}{\frac{\pi^{2}}{30}g_{\ast}T_{\ast}^{4}}$. $T_{\ast}$ is the first-order phase transition temperature. As discussed in Refs. \cite{Ahmadvand:2017xrw,Ahmadvand:2017tue}, we take $T_{\ast} \simeq T_{c}$ ($T_{c}$ is the critical temperature) and $g_{\ast}\sim 10$ at the phase transition temperature.

The latent heat $\epsilon_{\ast}$ is obtained as
\begin{equation}\label{i1}
\epsilon_{\ast}= \bigg(-\Delta F(T)+T \frac{d \Delta F(T)}{dT}\bigg)\bigg|_{T=T_{\ast}},
\end{equation}
where $\Delta F$ denotes the free energy differences.

\begin{figure}[H]
    \centering
      \setlength{\abovecaptionskip}{0.1cm}
    \includegraphics[width=10cm]{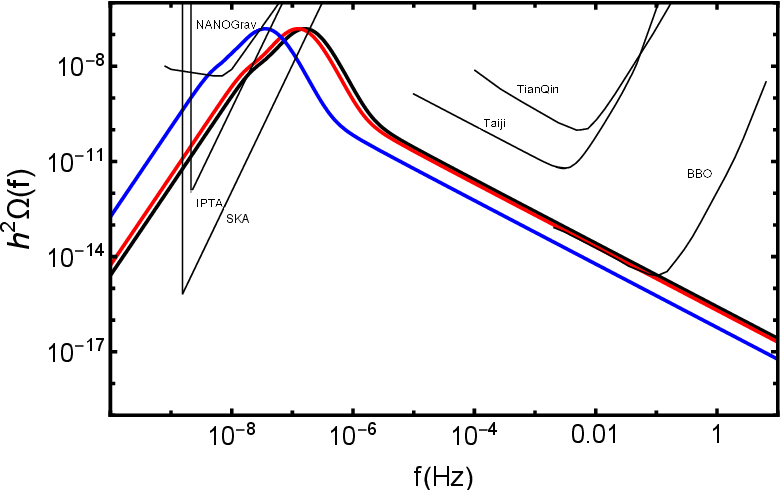}
    \caption{\label{fig1}  The GWs spectrum produced from the first order QCD phase transition with magnetic field for Jouguet detonations within hard wall model. From right to left $B=0.1, 0.4,$ and $0.7$, respectively. }
\end{figure}

The magnetized AdS black hole solution provides a fully backreacted and static magnetic field background, which lets us compute the equilibrium thermodynamics of the QCD phase transition. This is a separate and well-defined problem. Once these thermodynamic parameters are obtained, the subsequent GW production from MHD turbulence is a universal process that depends on these parameters ($\tau/H_*$ and  $v_b$), not on the detailed origin of the magnetic field. Therefore, it is completely standard to take the well-tested MHD turbulence GW spectrum from Ref.~\cite{Caprini:2009yp} with the parameters determined from our holographic calculation.

In our holographic setup, the magnetic field $B$ is a bulk quantity that serves as a theoretical input parameter, not a direct representation of the physical primordial magnetic field (PMF) strength in the early Universe. The values we consider ($B = 0.1$, $0.4$, $0.7$) lie well within the perturbative validity bound derived in Ref.~\cite{Dudal:2015wfn}. The magnetic field used in this paper are compatible with CMB and BBN constraints, our much larger values are not in conflict with those bounds. This is because our $B$ is a holographic parameter that controls the backreaction of the magnetic field on the geometry, rather than a direct physical field strength. As such, it is not subject to the same observational constraints. Our primary goal is to study the qualitative effect of such a magnetic field on the gravitational wave spectrum, not to claim a specific physical PMF strength.

The effects of magnetic helicity, inverse cascade, temperature-dependent viscosity and diffusivity, damping scales, finite source duration, unequal-time correlators, Hubble expansion, and a relativistic equation of state are explicitly incorporated in the standard MHD turbulence GW framework of Refs.~\cite{Caprini:2009yp,Caprini:2015zlo}. In particular, unequal-time correlators are treated with a top-hat approximation, turbulence is assumed to end when $\mathrm{Re}(L(T_{\mathrm{fin}})) = 3^{4/3}$, and the GW equation is solved in conformal time using $p = \rho/3$. Our calculation employs exactly the same framework.

We assume that the dominant effect of the magnetic field comes through its backreaction on the geometry, which alters the equilibrium thermodynamics. Corrections to the microphysical transport in MHD turbulence are taken to be subleading. Hence, our results should be interpreted as indicating qualitative trends. The MHD turbulence model we employed is exactly the one developed in Ref.~\cite{Caprini:2009yp} and subsequently used as the standard in Ref.~\cite{Caprini:2015zlo}. This model already includes cosmic expansion, finite source duration, unequal-time correlators, magnetic helicity, viscosity, and conductivity in a self-consistent analytical way. We readily admit that a full dynamical description of PMFs in holography is still an open problem. Nevertheless, our approach follows the best available practice in the literature.

\begin{figure}[H]
    \centering
      \setlength{\abovecaptionskip}{0.1cm}
    \includegraphics[width=10cm]{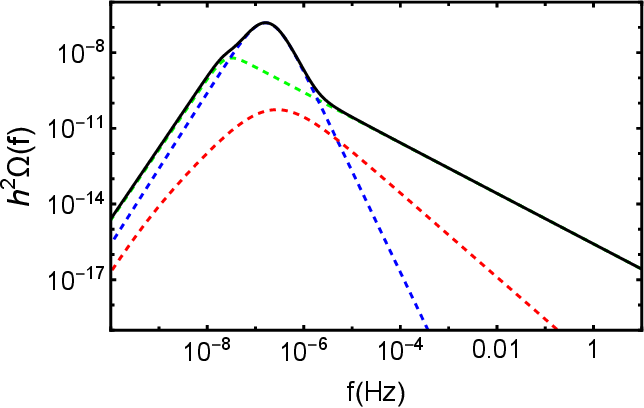}
    \caption{\label{fig2}  The total GWs frequency spectrum produced from the first order QCD phase transition at $B=0.1$ for Jouguet detonations within hard wall model. The solid black line denotes the total frequency spectrum. The contribution of bubble collision (dashed green line), sound waves (dashed blue line) and MHD turbulence (dashed red line) are plotted in this figure respectively. }
\end{figure}

\begin{figure}[H]
    \centering
      \setlength{\abovecaptionskip}{0.1cm}
    \includegraphics[width=10cm]{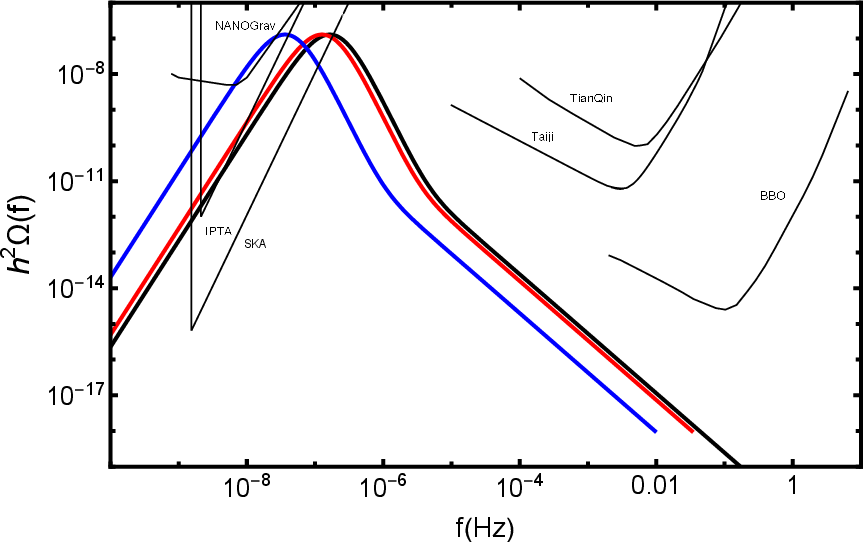}
    \caption{\label{fig3}  The GWs spectrum produced from the first order QCD phase transition with magnetic field for non-runaway bubbles within hard wall model. From right to left $B=0.1, 0.4,$ and $0.7$, respectively. }
\end{figure}

\begin{figure}[H]
    \centering
      \setlength{\abovecaptionskip}{0.1cm}
    \includegraphics[width=10cm]{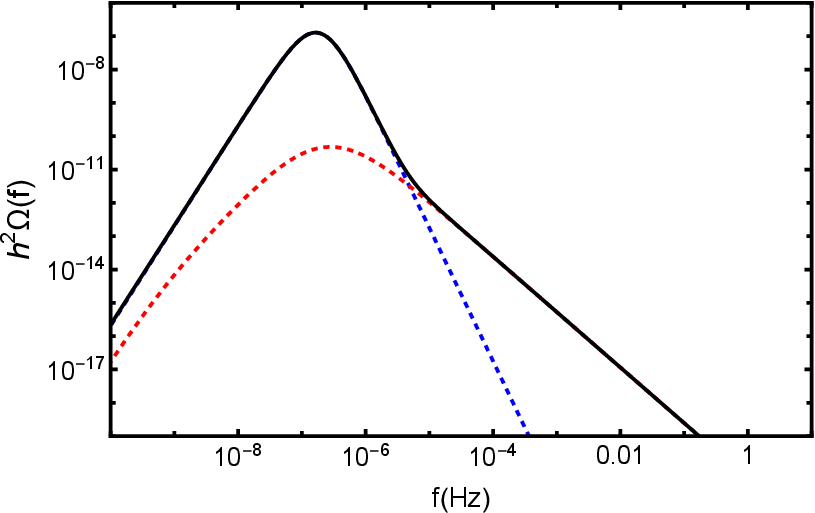}
    \caption{\label{fig4}  The total GWs frequency spectrum produced from the first order QCD phase transition at $B=0.1$ for non-runaway bubbles within hard wall model. The solid black line denotes the total frequency spectrum. The contribution of sound waves (dashed blue line) and MHD turbulence (dashed red line) are plotted in this figure respectively. }
\end{figure}

\begin{figure}[H]
    \centering
      \setlength{\abovecaptionskip}{0.1cm}
    \includegraphics[width=10cm]{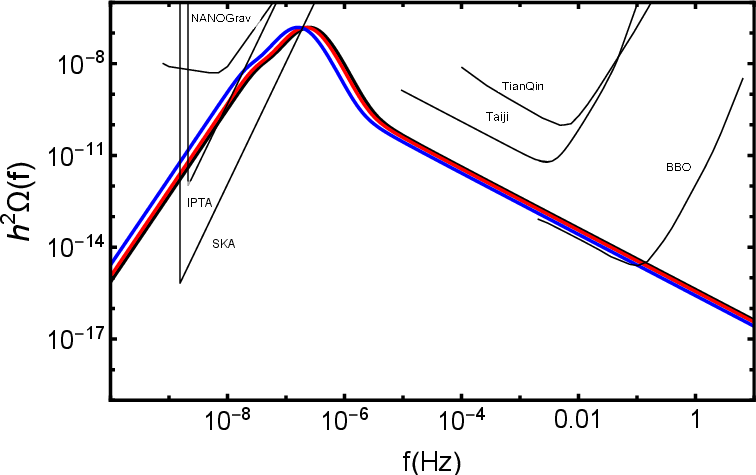}
    \caption{\label{fig5}  The GWs spectrum produced from the first order QCD phase transition with magnetic field for Jouguet detonations within soft wall model. From right to left $B=0.1, 0.4,$ and $0.7$, respectively. }
\end{figure}

\begin{figure}[H]
    \centering
      \setlength{\abovecaptionskip}{0.1cm}
    \includegraphics[width=10cm]{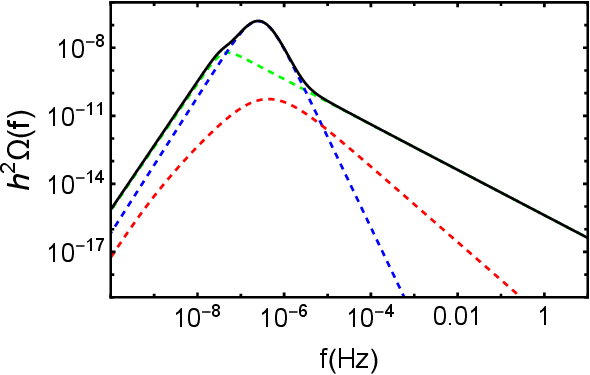}
    \caption{\label{fig6}  The total GWs frequency spectrum produced from the first order QCD phase transition at $B=0.1$ for Jouguet detonations within soft wall model. The solid black line denotes the total frequency spectrum. The contribution of bubble collision (dashed green line), sound waves (dashed blue line) and MHD turbulence (dashed red line) are plotted in this figure respectively. }
\end{figure}

\begin{figure}[H]
    \centering
      \setlength{\abovecaptionskip}{0.1cm}
    \includegraphics[width=10cm]{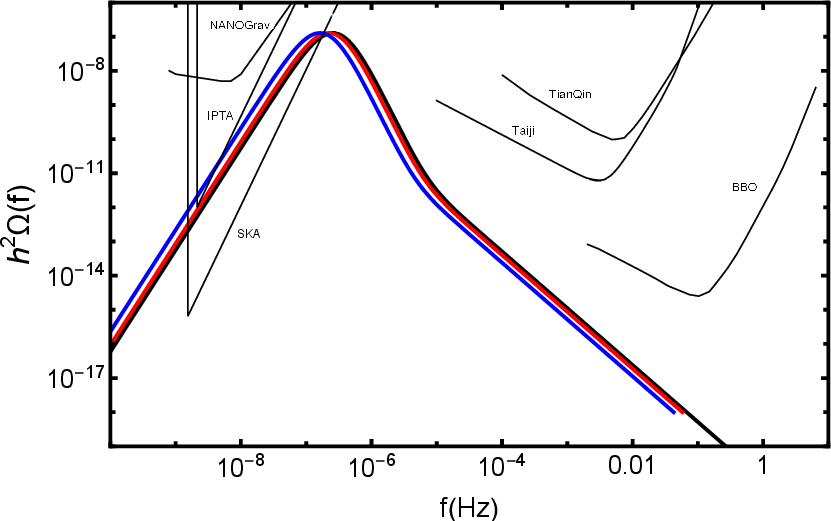}
    \caption{\label{fig7} The GWs spectrum produced from the first order QCD phase transition with magnetic field for non-runaway bubbles within soft wall model. From right to left $B=0.1, 0.4,$ and $0.7$, respectively. }
\end{figure}

\begin{figure}[H]
    \centering
      \setlength{\abovecaptionskip}{0.1cm}
    \includegraphics[width=10cm]{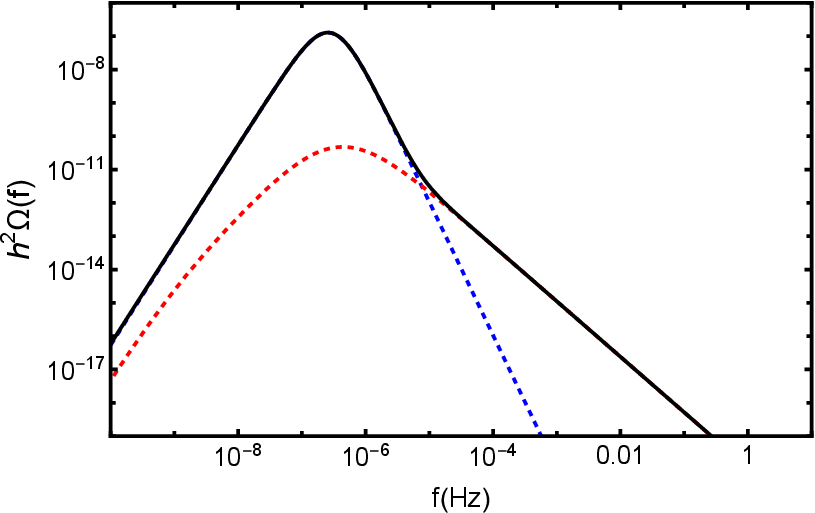}
    \caption{\label{fig8}  The total GWs frequency spectrum produced from the first order QCD phase transition at $B=0.1$ for non-runaway bubbles within soft wall model. The solid black line denotes the total frequency spectrum. The contribution of sound waves (dashed blue line) and MHD turbulence (dashed red line) are plotted in this figure respectively. }
\end{figure}

In the calculations, we take $\tau/H_{\ast}=10$ as discussed in Refs. ~\cite{Ahmadvand:2017xrw,Ahmadvand:2017tue,Chen:2017cyc}. The effects of magnetic field on the GWs spectrum are shown for the hard wall model in Figs. \ref{fig1}-\ref{fig4}, and for the soft wall model in Figs. \ref{fig5}-\ref{fig8}.

Fig.~\ref{fig1} displays the effect of magnetic field on the total GWs spectrum for Jouguet detonations in hard wall model. This figure shows that the position of peak frequency moves backward as magnetic field increases. Moreover, the total GWs spectrum could be detected by the International Pulsar Timing Array (IPTA), Square Kilometre Array (SKA), Big-Bang Observer (BBO) and the North American NanoHertz Observatory for Gravitational Waves (NANOGrav) \cite{Moore:2014lga,NANOGrav:2020bcs}, while might not be detected by the TianQin and Taiji \cite{Ruan:2018tsw}.

Fig. \ref{fig2} shows the individual contributions of sound waves, bubble collisions, and MHD turbulence to the total gravitational waves spectrum for Jouguet detonations in hard wall model. The spectrum from bubble collisions aligns with the total spectrum in the frequency bands below \(2.8 \times 10^{-8}\) Hz and above \(2.6 \times 10^{-6}\) Hz, indicating that bubble collisions dominate the GWs signal in these regions. Sound waves contribute primarily in the intermediate frequency range between \(2.8 \times 10^{-8}\) Hz and \(2.6 \times 10^{-6}\) Hz, with the spectral peak located around \(1.5 \times 10^{-7}\) Hz. In contrast, the contribution from MHD turbulence to the total GWs spectrum is negligible across the studied frequency range.

Fig. \ref{fig3} presents the magnetic field dependence of the total GWs spectrum for non-runaway bubbles within the hard wall model. The results show that the spectral peak shifts to lower frequencies as the magnetic field increases. In terms of detectability, this spectrum lies within the sensitivity ranges of IPTA, SKA, and NANOGrav, but falls outside those of BBO, TianQin, and Taiji. Fig. \ref{fig4} displays the contributions from sound waves and MHD turbulence to the total GW spectrum for non-runaway bubbles in the hard wall model. Sound waves dominate the signal at frequencies below \(6.4 \times 10^{-6}\) Hz, producing a spectral peak near \(1.7 \times 10^{-7}\) Hz. Above this frequency threshold, the contribution from MHD turbulence becomes predominant.

Fig. \ref{fig5} shows how the magnetic field affects the total gravitational wave spectrum for Jouguet detonations in the soft wall model. The spectral peak shifts slightly toward lower frequencies as the magnetic field increases. The predicted signal is potentially detectable by IPTA, SKA, and BBO, but likely lies outside the sensitivity ranges of NANOGrav, TianQin, and Taiji. Fig. \ref{fig6} shows the contributions to the total GWs spectrum from sound waves, bubble collisions, and MHD turbulence for Jouguet detonations within soft wall model. Bubble collisions dominate at both low (\(< 3.8 \times 10^{-8}\) Hz) and high (\(> 4.5 \times 10^{-6}\) Hz) frequencies. Sound waves are the main contributor in the intermediate range (\(3.8 \times 10^{-8}\) Hz to \(4.5 \times 10^{-6}\) Hz) and produce the spectral peak near \(2.7 \times 10^{-7}\) Hz. MHD turbulence remains negligible across the entire frequency range.

Fig. \ref{fig7} shows how the total gravitational wave spectrum for non-runaway bubbles in the soft wall model depends on the magnetic field. A stronger magnetic field slightly shifts the spectral peak toward lower frequencies. This signal may be detectable by IPTA and SKA, but likely not by BBO, NANOGrav, TianQin, or Taiji. Fig. \ref{fig8} displays the contributions from sound waves and MHD turbulence in this scenario. Sound waves dominate the GW spectrum below \(1.5 \times 10^{-5}\) Hz, generating a peak at about \(2.95 \times 10^{-7}\) Hz. Above this threshold, MHD turbulence becomes the main contributor.

Our qualitative conclusion is that a stronger magnetic field shifts the peak of the gravitational wave spectrum to lower frequencies. This suggests that the direction of the peak shift is a robust feature, not just some artifact of a particular model. The authors of Ref.~\cite{Dudal:2015wfn} show that in both hard-wall and soft-wall models, the critical deconfinement temperature $T_c$ drops as the magnetic field $B$ increases. Since the peak frequency of gravitational waves from a first-order phase transition basically scales with the transition temperature, that shift with $B$ follows naturally from this shared thermodynamic behavior. Thus, the direction of the shift is a qualitative trend independent of the specific model.

We fully acknowledge that the calculation of the bubble nucleation rate in holographic QCD is non-trivial It would require solving the Euclidean action in the five-dimensional background or building an effective four-dimensional dilaton potential \cite{Agashe:2019lhy}. We simply take $\tau/H_* = 10$, a value widely used in Refs.~\cite{Ahmadvand:2017xrw,Ahmadvand:2017tue,Chen:2017cyc}. Even the latest state-of-the-art holographic models still treat $\tau/H_*$ as a free parameter (with a bound $\tau/H_* > 8.59$) \cite{He:2022amv}. The work of Ref.\cite{He:2022amv} shows computing the nucleation rate from first principles is still an open challenge, and the field typically adopts either typical values or treats the duration as a free parameter.

Most importantly, our qualitative conclusion does not rely on the exact value of $\tau/H_*$. From the scaling $f_{\text{peak}} \propto (\tau/H_*) T_*$ in Eq.~(17), the ratio for two different magnetic fields becomes $\frac{f_{\text{peak}}(B_2)}{f_{\text{peak}}(B_1)} \approx \frac{T_*(B_2)}{T_*(B_1)}$, since $\tau/H_*$ drops out as long as its dependence on $B$ is subleading.Since the dominant effects of the magnetic field are already captured by the thermodynamic quantities we compute from the magnetized Hawking-Page transition, taking $\tau/H_*$ as constant is the simplest and most natural choice. The predicted shift of the peak is therefore directly controlled by $T_*(B)$ and stays robust regardless of uncertainties in the nucleation rate. A more refined nucleation treatment in magnetized holography is certainly an important direction for future work.

\section{Conclusion and discussion}\label{sec:04}

In this work, we investigate the gravitational waves produced by a first-order QCD confinement-deconfinement phase transition in the presence of a magnetic field from holography. We calculate the GWs spectra for both the hard wall and soft wall models, considering Jouguet detonations and non-runaway scenarios.

A key finding is that an increasing magnetic field consistently shifts the peak frequency of the total GWs spectrum to lower values across all models and scenarios. Regarding detectability, the predicted spectra from both models are accessible to IPTA and SKA. Under certain conditions, signals may also be detectable by BBO or NANOGrav. In most cases, the peak of the spectrum is generated by sound waves, which dominate the signal within a specific intermediate frequency range. Bubble collisions typically provide the leading contribution at both the lowest and highest frequencies considered. Notably, the contribution from MHD turbulence to the total spectrum is found to be negligible in the Jouguet detonation scenarios for both holographic models. However, for non-runaway bubbles, MHD turbulence becomes the predominant source at higher frequencies, above a specific threshold.

We agree that a realistic assessment of detectability with PTA requires careful modeling of various foregrounds and noise sources, including supermassive black hole binaries (SMBHBs), red noise, clock noise, and other astrophysical and cosmological backgrounds. Our comparison with PTA sensitivity curves is only meant as a first-order, qualitative illustration. This is actually quite common in the literature on gravitational waves from cosmological phase transitions Refs.~\cite{Caprini:2015zlo,Ahmadvand:2017xrw}. We simply want to show that the predicted peak frequency and amplitude land in a range where future PTAs could be sensitive. This is the usual way to identify promising parameter regions before moving on to full Bayesian analyses.

More importantly, a recent Bayesian analysis by Ref.~\cite{He:2023ado} used the NANOGrav 15-year data to constrain a holographic QCD model (without a magnetic field). Their results indicate that first-order QCD phase transitions with $\tau/H_* \sim 1.3\text{--}2.5$ are compatible with PTA data. Our adopted value $\tau/H_* = 10$ is a few times larger, yet our conclusion that the GW peak shifts to lower frequencies with increasing $B$ does not depend on its exact value. So even if the true $\tau/H_*$ turned out to be closer to the value inferred by Ref.~\cite{He:2023ado}, our prediction that the magnetic field shifts the peak to lower frequencies remains unchanged.

We certainly agree that a full Bayesian model comparison, including SMBHB backgrounds, cosmic strings, inflationary relics, and other phase transition models, is necessary to identify the true origin of the PTA signal. Such an analysis is beyond the scope of the present work, which focuses on the qualitative effect of a magnetic field on the GW spectrum.

Our analysis suggests that a first-order QCD phase transition in a magnetized early universe remains a viable cosmological source for GWs signal potentially observable by current and upcoming pulsar timing arrays. The characteristic spectral features provide a potential observational signature. Future, more sensitive PTA observations could thus serve as a powerful tool to probe the conditions of the primordial plasma, potentially constraining the strength and cosmological impact of primordial magnetic fields.

\section*{Acknowledgments}

%\begin{acknowledgements}
Manman Sun is supported by the National Natural Science Foundation of China under Grant No. 12305076. Zhou-Run Zhu is supported by the Natural Science Foundation of Henan Province of China under Grant No. 242300420947. Zhou-Run Zhu is also supported by the High Level Talents Research and Startup Foundation Projects for Doctors of Zhoukou Normal University No. ZKNUC2023018.
%\end{acknowledgements}

\end{document}